# QZ Carinae—Orbit of the Two Binary Pairs

**Mark Blackford**
*Variable Stars South (VSS), Congarinni Observatory, Congarinni, NSW, Australia 2447; markgblackford@outlook.com*

**Stan Walker**
*Variable Stars South (VSS), Wharemaru Observatory, Waiharara, Northland, New Zealand 0486*

**Edwin Budding**
*Variable Stars South (VSS), Carter Observatory, Kelburn, Wellington, New Zealand 6012*

**Greg Bolt**
*Variable Stars South (VSS), Craigie Observatory, Craigie, WA, Australia 6025*

**Dave Blane**
*Variable Stars South (VSS), and Astronomical Society of Southern Africa (ASSA), Henley Observatory, Henley on Klip, Gauteng, South Africa*

**Terry Bohlsen**
*Variable Stars South (VSS), and Southern Astro Spectroscopy Email Ring (SASER), Mirranook Observatory, Armidale, NSW, Australia, 2350*

**Anthony Moffat**
*BRITE Team, Département de physique, Université de Montréal, CP 6128, Succursale Centre-Ville, Montréal, QC H3C 3J7, Canada*

**Herbert Pablo**
*BRITE Team, American Association of Variable Star Observers, 49 Bay State Road, Cambridge, MA 02138*

**Andrzej Pigulski**
*BRITE Team, Instytut Astronomiczny, Uniwersytet Wrocławski, Wrocław, Poland*

**Adam Popowicz**
*BRITE Team, Department of Automatic Control, Electronics and Informatics, Silesian University of Technology, Gliwice, Poland*

**Gregg Wade**
*BRITE Team, Department of Physics and Space Science, Royal Military College of Canada, P.O. Box 17000, Station Forces, Kingston, ON K7K 7B4, Canada*

**Konstanze Zwintz**
*BRITE Team, Universität Innsbruck, Institut für Astro- und Teilchenphysik, Technikerstrasse 25, A-6020 Innsbruck, Austria*

**Abstract**   We present an updated O–C diagram of the light-time variations of the eclipsing binary (component B) in the system QZ Carinae as it moves in the long-period orbit around the non-eclipsing pair (component A). This includes new Variable Stars South members' measures from 2017 to 2019, BRITE satellite observations in 2017 and 2018, and 100 previously unpublished measures made at Auckland Observatory from 1974 to 1978. We conclude that QZ Carinae has not yet completed one orbit of the two pairs since discovery in 1971. The duration of totality of primary eclipses was measured to be $0.295 \pm 0.02$ day ($7.08 \pm 0.48$ hours), rather longer than earlier values from light curve models. Other observational findings include the shape of primary and secondary eclipses and small-scale short-term brightness changes.

### 1. Introduction

QZ Carinae (= HD 93206, HIP 52526; V = 6.24, U–B = –0.84, B–V = 0.13; Wenger *et al.* 2000) is the brightest member of the open cluster Collinder 228 within the Great Carina Nebula region. Variability was first discovered by Brian Marino and Stan Walker at the Auckland Observatory in 1971 (Walker and Marino 1972).

Spectroscopy by Morrison and Conti (1979) revealed the system to comprise at least two pairs of very massive stars. The primary pair, non-eclipsing, has an orbital period of ~20.73 days. The less luminous eclipsing pair has a period of 5.99857 days. Leung *et al.* (1979) provided further details of these four stars, giving a total mass of 93 ± 12.6 solar masses. They also derived a mutual orbital period of several decades for the two pairs orbiting around each other.

The model arising from these early investigations is essentially similar to that of present-day understanding. According to Parkin *et al.* (2011) the brightest component of the combined spectrum is the O9.7 I type supergiant in "Component A." The secondary in Component A has not been directly observed but it is thought to be an early B-type dwarf (B2 V). "Component B" is the eclipsing binary whose brighter member is the less massive O8 III type star. Its eclipsing companion is presumed to be a more massive late O-type dwarf (O9 V). Spectral signs of this star have been mentioned in the literature, but clear evidence has been difficult to demonstrate and indications of additional variability in the system complicate the picture.

The "eclipse method" is well established as a source of empirical knowledge of stellar properties (cf. e.g. Eker *et al.* 2018). As one of the most massive close systems known, QZ Car has a special role in informing about these properties at the high end of the stellar mass range. Such young hot stars are associated with strong radiation fields and stellar winds that interact with powerful shock fronts giving rise to significant X-ray emission (Parkin *et al.* 2011). Given this context, it could be reasonably proposed that QZ Car should be an ancestor of a future gravitational wave source. This observationally challenging system may be the best test-case for checking the Roche-lobe radiative-distortion effects predicted by Drechsel *et al.* (1995).

At face value, Component B's arrangement is strongly suggestive of Case A (hydrogen core burning mass donor) type interactive binary evolution. Morrison and Conti (1980) mentioned that speckle interferometry could separate components A and B and help clarify their physical parameters. The O–C diagram from eclipse timings should also bear on this. If the full facts on a multiple stellar system such as QZ Car were available, it would offer critical tests of stellar evolution modelling, including the relationship of stars to their Galactic environment (Andersen *et al.* 1993).

In a more recent review of the system, Walker *et al.* (2017) suggested higher masses than Leung *et al.* (1979), as shown in their Tables 9 and 10 based on a value of 49.5 years for the long-period orbit. The exact period would constrain the total masses of the individual binaries and thence establish more precise parameters of the system components.

The present paper concentrates on our knowledge of the long-period orbit.

## 2. Ground-based observations

Ground-based measures were compiled from a variety of sources, some not previously published. Until about 1995 measures were single-channel differential photoelectric photometry, usually in UBV. Since then they were almost entirely CCD in B and V, except for some with DSLR cameras. Only the V magnitudes have been used in this study.

Due to the lack of uniformity of comparison stars in the photoelectric era no attempt has been made to correct published magnitudes. Since 1995 most measures have been linked to E Region standards although a few used the AAVSO's APASS system which has some slight divergence.

A variety of comparison stars and telescopes have been used leading to zero point offsets between the different sets of observations. This does not affect our conclusions. The unpublished measures were made in UBV at Auckland Observatory as part of a project monitoring eta Carinae. Exactly 100 in number, they extend from 1974 to 1978 and allow the determination of six new epochs which fill in the lower part of the O–C diagram very well.

Between 2017 and 2019 time series observations were collected from two sites to better define the duration of totality of primary eclipses. GB observed from the west coast of Australia (115.75833° E, 31.78917° S) through a V filter and 25-cm telescope stopped down to 8 cm. MB observed from the east coast of Australia (152.86040° E, 30.73452° S) through a V filter and 8-cm refractor stopped down to 5 cm. For both setups exposure times were limited to 10 seconds or less to avoid saturation. To reduce the effect of scintillation 5 to 7 observations were averaged to obtain the final magnitudes.

## 3. BRITE satellite observations

The BRIght Target Explorer (BRITE) nano-satellites (Weiss *et al.* 2014; Pablo *et al.* 2016) provide high precision photometry of bright stars (typically V < 5 mag) for continuous periods of up to 180 days. The un-cooled CCD cameras have an effective field of view of 24 × 20 degrees, however only a limited number of small sub-rasters, centered on pre-selected stars of interest, are downloaded for each pointing of a satellite.

BRITE images are deliberately defocussed to avoid undersampling due to the 27-arcsec/pixel resolution. In crowded regions, such as the QZ Car field, blending of multiple star images is unavoidable. Furthermore, radiation-induced defects in the sensors and other instrumental issues complicate image acquisition and analysis processes (Popowicz *et al.* 2017; Pigulski *et al.* 2018).

Although not originally selected as a target, QZ Car was included in the raster of the Wolf-Rayet star WR24 (HD 93131) which was observed between 2017 Jan 29 and Jul 1 by the red-filter BRITE-Heweliusz (BHr) satellite. Data for WR24 and QZ Car were successfully extracted and de-trended separately.

A total of 77,7783 observations were taken in chopping mode (Pablo *et al.* 2016) with exposure times of 5 seconds. Typically about 50 images were recorded over approximately 30 minutes during each 97.0972-minute orbit of the BHr satellite. QZ Car does not vary significantly on such short time scales so to improve the signal-to-noise ratio we used the orbit-averaged magnitudes (see Figure 1). The light curve phased on a period of 5.99857 d is shown in Figure 2.

QZ Car was observed again by BRITE-Heweliusz between 2018 May 16 and Jul 1, and by the red-filter BRITE-Toronto (BTr) satellite between 2018 Feb 16 and May 19.

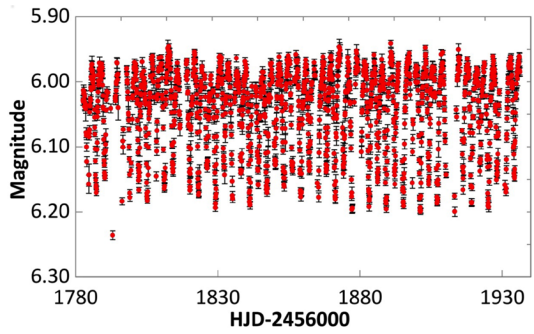

Figure 1. BRITE-Heweliusz 2017 QZ Car light curve, observations detrended and averaged per spacecraft orbit.

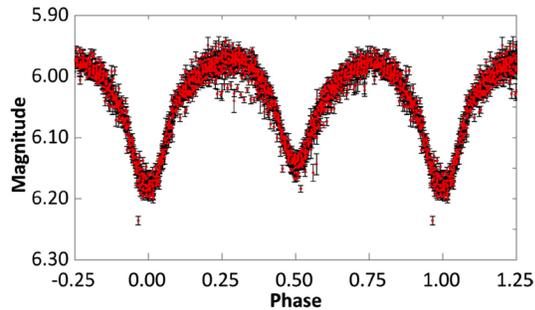

Figure 2. BRITE-Heweliusz 2017 phased light curve based on a period of 5.99857 d.

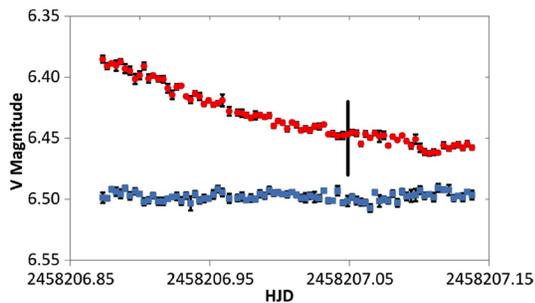

Figure 3. Ground-based light curves of QZ Car (red) and the check star HD 93191 (blue) with their error bars (black). The check star was shifted by 2 magnitudes for display purposes. Predicted start of totality based on the Leung *et al.* 1979 model is indicated by the vertical black line.

### 4. Determination of epochs of eclipse minima

The long-term near-continuous BRITE satellite observations allowed precise epochs of eclipse minimum to be easily measured. Determining epochs of eclipse minimum is also straightforward from ground-based observations when both ingress and egress are recorded in a single observing session. However, the long duration of totality (~5.76 hours as modelled by Leung *et al.* 1979) and small amplitude (~0.24 in V due to the more luminous non-eclipsing pair) make this difficult to apply to ground-based observations of QZ Car.

At 60° S declination QZ Car is accessible only from the southern hemisphere where experienced observers in suitable locations are relatively few. The magnitude of the star is too bright for most CCD observers. All recent CCD and DSLR measures have been made with stopped down telescopes and/or short exposures.

From latitude 30° S the system is at opposition in early

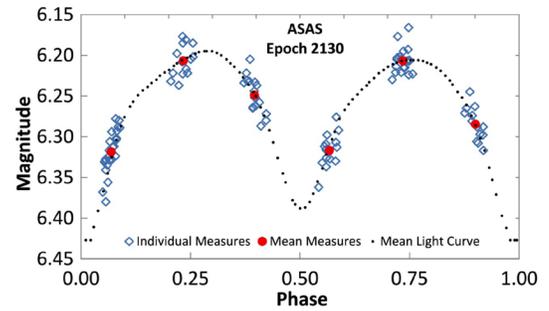

Figure 4. BRITE mean light curve (black dotted line) with 101 individual ASAS measures (open blue diamonds) between 2005 Oct and 2006 Jun. Filled red circles are average values within each group of observations.

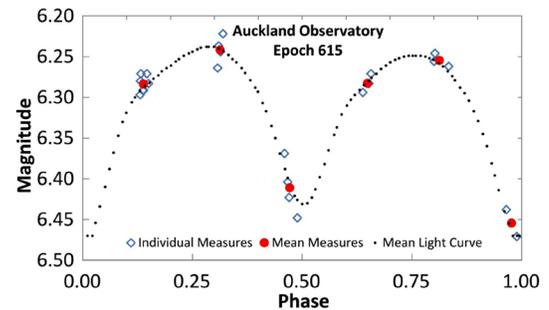

Figure 5. Similar to Figure 4 but with only 23 measures which, when averaged, fit the mean light curve well enough to determine a reliable epoch.

March when astronomical darkness is ~10 hours, leaving only 2 hours before and after totality to measure parts of ingress and egress. The magnitude change over that time frame is very small, of the order of a few hundredths of a magnitude. Thus, second and third contact points defining the duration of totality are hard to determine accurately. An example is shown in Figure 3. Measured times of the second contact point, when ingress changes to totality, have abnormally large measurement error due to the light curve slope during totality. It was not possible to observe the end of totality in this case.

The period of 5.99857 days is also inconvenient. Only a narrow longitude range is suitable for observing the entire period of totality at opposition. From a given observing location eclipses occur 0.00143 d earlier each cycle, or 0.0817 d per year. So the ideal longitude moves east by 31.3° per year, taking 11.5 years to complete the cycle, although due to the light curve shape this time can be halved.

We adopted the method of fitting a mean light curve to random measures of QZ Car during annual seasons. This produces good annual epochs and is often used with longer period Cepheids. Initially we employed the symmetrical mean light curve determined by Leung *et al.* (1979) from the original observations of Walker and Marino (1972). However, Hipparcos (Perryman *et al.* 1997) and later ground-based observations indicated that the light curve is asymmetric. In particular, maxima following primary eclipses are significantly brighter than maxima following secondary eclipses. Unfortunately, a reliable mean light curve could not be constructed from the rather noisy and incomplete Hipparcos and ground-based measures.

We resolved this by constructing a mean light curve of 2017 BRITE measures. The 0.203-magnitude range of this mean light curve is smaller than the 0.230-magnitude range measured in the V band. This results from blending with nearby stars and the non-standard red filter employed in BRITE-Heweliusz. The mean light curve was therefore scaled to match the V magnitude range and a zero point correction applied to account for different comparison stars used by each observer. Epochs derived are not affected by the zero point corrections.

The BRITE mean light curve is shown in Figure 4 along with 101 measures from one season of the All Sky Automated Survey (Pojmański 1997). It can also be used with fewer random measures as shown in Figure 5. This method provides a visual picture of the accuracy which is very useful. Much of the scatter in individual observations can be attributed to small-scale brightness variations discussed in section 7. The effect of these variations can be minimized by averaging values within each group of observations.

## 5. Observed-Minus-Calculated diagram of epochs of primary minimum

Mayer *et al.* (2001) presented an O–C diagram of six published times of minimum for QZ Car from discovery in 1971 to 1994. The light-time effect (LTE) due to the mutual orbit of the two pairs around each other was evident. Despite only a fraction of the orbit being covered they were able to conclude that the period is several decades.

We compiled a comprehensive set of published, historic (but previously unpublished) and new observations of QZ Car. Epochs of primary eclipses were derived from seasonal measurements (Table 1) and a more complete O–C diagram constructed in an attempt to better determine the period of the mutual orbit of the two pairs. In Figure 6 we show the O–C diagram covering 48 years since 1971 using the light elements JD $2441033.033 + 5.99857 \times E$, where E is the number of epochs since the initial epoch.

There are two related uncertainties in this area. Until the value of the long period orbit is determined the period derived for the eclipsing pair by Mayer *et al.* (1998) is uncertain at the fifth decimal level. This does not affect seasonal epochs. The period of the eclipsing pair will be known when one long-period orbit is completed—but the long period (and hence the correction to the eclipse period) cannot be determined until the O–C curve begins to duplicate itself.

The completion of one cycle of the long-period orbit will be seen when the O–C curve begins to run parallel to the early cycles. The period will be uncertain until the rate of approach begins to slow and the O–C curve becomes flatter.

The slope of the O–C diagram was 5.99818 days/cycle between cycles 0 and 185 and 5.99833 days/cycle between cycles 2676 and 2933. Both are shorter than the adopted period of 5.99857 days as the eclipsing system was approaching us in both cases. The period between cycles 185 and 433 had lengthened to 5.998454 days, clearly showing the curvature of the long orbit as it nears the closest point to the observer.

It is likely that the current observed trend will continue for several more years before following the curve shown from cycle 700 onward where it begins a change to the longer receding period.

## 6. Determining the duration of totality

### 6.1. Primary eclipses

For the 2019 season we organized a wider spread in longitude in an attempt to observe the two points defining the beginning and end of totality. These were made using aperture masks and V filters by GB at Craigie Observatory (115.75833° E) and MB at Congarinni Observatory (152.86040° E). Unfortunately, they were not able to observe on the same nights. Together with measures by MB in 2017 and 2018 we determined 8 ingress points and 4 egress points.

Figure 7 is a composite graph of primary eclipse totality. Measures from MB (cycle 2789) during ingress and up to mid-totality are combined with those of GB (cycle 2925) during totality and egress. GB's measures were corrected for the change in LTE between the two cycles. From these we determined duration of totality to be $0.295 \pm 0.02$ day. Totality is not flat, which complicates the determination of second and third contact times, although end of totality was generally clearer. This illustrates why early measures from single sites were ineffective.

Figure 8 shows the O–C values of our measured seasonal epochs and primary eclipse second and third contact points defining the start and end of totality. The scatter in the beginning and ending points illustrates the difficulty in determining exact times due to the low amplitude of eclipses.

Measures during the past three years indicate that totality lasts $7.08 \pm 0.48$ hours, rather longer than the 5.76 hours of the original Leung *et al.* (1979) model but a better fit to the wider eclipses of Walker *et al.* (2017).

### 6.2. Secondary eclipses

The most complete measures available of a secondary eclipse from one site are those of Grant Christie from Auckland Observatory in 1994 (Mayer *et al.* 1998) but these do not cover totality in full (Figure 9). Also shown is a composite light curve from ingress measures by MB in 2018 and egress measures by GB in 2019.

The considerable curvature is due to this eclipse being an annular transit by the smaller star. We were unable to identify second and third contact points of these annular transits from such light curves.

BRITE satellite data were unsuitable for determining start and end of totality. Imaging was restricted to about 30 minutes within each satellite orbit and individual measures had relatively large uncertainties. Orbit-averaged magnitudes, while more precise, had relatively poor time resolution.

## 7. Other brightness variations

Larger-than-normal scatter in seasonal light curves (cf. Figure 4) indicates variation over and above that due to the eclipsing pair. Hipparcos measures also showed considerable scatter (Figure 1 of Walker *et al.* 2017). The masses and spectral types of the four stars making up the system, in particular the O9 type supergiant, are such that slight variations in brightness are not unexpected. Some supergiants show relatively strong variability from surface bright spots (cf. Ramiaramanantsoa *et al.* 2018).

Table 1. Derived seasonal epochs of central primary eclipse. Values in column 2 calculated using zero epoch 41033.033 and period 5.99857 days. All measures used in this paper may be accessed at: ftp://ftp.aavso.org/public/datasets/200112_QZ_Car_Observations.xlsx

| Cycle | Calculate (HJD – 2400000) | Derived (HJD – 2400000) | LTE (d) | error (d) | Number of Observations | Source (Observer) |
|---|---|---|---|---|---|---|
| 0 | 41033.033 | 41033.033 | 0.000 | 0.015 | 41 | Auckland Photoelectric Observers' Group |
| 21 | 41159.003 | 41159.002 | –0.001 | 0.015 | 41 | Auckland Photoelectric Observers' Group |
| 46 | 41308.967 | 41308.953 | –0.014 | 0.019 | 17 | Auckland Photoelectric Observers' Group |
| 133 | 41830.843 | 41830.793 | –0.050 | 0.045 | 12 | Auckland Photoelectric Observers' Group |
| 185 | 42142.768 | 42142.699 | –0.069 | 0.018 | 53 | Auckland Photoelectric Observers' Group |
| 264 | 42616.655 | 42616.586 | –0.070 | 0.034 | 12 | Auckland Photoelectric Observers' Group |
| 309 | 42886.591 | 42886.512 | –0.079 | 0.015 | 14 | Auckland Photoelectric Observers' Group |
| 368 | 43240.507 | 43240.418 | –0.089 | 0.022 | 5 | Auckland Photoelectric Observers' Group |
| 432 | 43624.415 | 43624.320 | –0.095 | 0.022 | 15 | Auckland Photoelectric Observers' Group |
| 555 | 44362.239 | 44362.239 | 0.000 | 0.033 | 13 | Auckland Photoelectric Observers' Group |
| 615 | 44722.154 | 44722.037 | –0.116 | 0.025 | 23 | Auckland Photoelectric Observers' Group |
| 672 | 45064.072 | 45063.962 | –0.110 | 0.028 | 8 | Auckland Photoelectric Observers' Group |
| 759 | 45585.948 | 45585.828 | –0.120 | 0.028 | 8 | Auckland Photoelectric Observers' Group |
| 1161 | 47997.373 | 47997.324 | –0.049 | 0.025 | 38 | Hipparcos |
| 1201 | 48237.316 | 48237.274 | –0.042 | 0.016 | 19 | Hipparcos |
| 1243 | 48489.256 | 48489.202 | –0.054 | 0.032 | 46 | Hipparcos |
| 1278 | 48699.205 | 48699.147 | –0.059 | 0.020 | 150 | Pavel Mayer (La Silla Observatory) |
| 1299 | 48825.175 | 48825.126 | –0.049 | 0.018 | 24 | Hipparcos |
| 1331 | 49017.130 | 49017.091 | –0.039 | 0.056 | 54 | Pavel Mayer (La Silla Observatory) |
| 1350 | 49131.103 | 49131.083 | –0.020 | 0.019 | 9 | Auckland Photoelectric Observers' Group |
| 1398 | 49419.034 | 49419.022 | –0.012 | 0.018 | 57 | Pavel Mayer (Christie) |
| 1398 | 49419.034 | 49419.034 | 0.000 | 0.038 | 20 | Auckland Photoelectric Observers' Group |
| 1948 | 52718.247 | 52718.517 | 0.270 | 0.040 | 21 | All Sky Automated Survey |
| 2006 | 53066.164 | 53066.444 | 0.280 | 0.010 | 56 | All Sky Automated Survey |
| 2015 | 53120.152 | 53120.442 | 0.290 | 0.031 | 43 | Pavel Mayer (SAAO) |
| 2074 | 53474.067 | 53474.368 | 0.301 | 0.013 | 77 | All Sky Automated Survey |
| 2129 | 53803.989 | 53804.305 | 0.316 | 0.008 | 101 | All Sky Automated Survey |
| 2196 | 54205.893 | 54206.190 | 0.297 | 0.019 | 90 | All Sky Automated Survey |
| 2253 | 54547.811 | 54548.133 | 0.322 | 0.014 | 108 | All Sky Automated Survey |
| 2309 | 54883.731 | 54884.071 | 0.340 | 0.017 | 72 | All Sky Automated Survey |
| 2324 | 54973.710 | 54974.040 | 0.330 | 0.040 | 26 | Variable Stars South (Bohlsen) |
| 2376 | 55285.635 | 55285.935 | 0.300 | 0.031 | 45 | Variable Stars South (Bohlsen) |
| 2562 | 56401.369 | 56401.639 | 0.270 | 0.018 | 36 | Variable Stars South (Bohlsen) |
| 2675 | 57079.208 | 57079.483 | 0.275 | 0.023 | 47 | AAVSO (BSM South) |
| 2676 | 57085.206 | 57085.476 | 0.270 | 0.018 | 122 | Variable Stars South (Blackford) |
| 2683 | 57127.196 | 57127.476 | 0.280 | 0.020 | 38 | AAVSO (BSM Berry) |
| 2735 | 57439.122 | 57439.372 | 0.250 | 0.021 | 69 | Variable Stars South (Blackford) |
| 2796 | 57805.035 | 57805.263 | 0.228 | 0.006 | 1681 | BRITE |
| 2799 | 57823.030 | 57823.280 | 0.250 | 0.013 | 121 | Variable Stars South (Bolt) |
| 2800 | 57829.029 | 57829.279 | 0.250 | 0.022 | 96 | Variable Stars South (Blane) |
| 2809 | 57883.016 | 57883.266 | 0.250 | 0.015 | 124 | Variable Stars South (Blane) |
| 2861 | 58194.942 | 58195.172 | 0.230 | 0.014 | 124 | Variable Stars South (Blackford) |
| 2863 | 58206.939 | 58207.169 | 0.230 | 0.008 | 100 | Variable Stars South (Blackford) |
| 2859 | 58182.945 | 58183.167 | 0.222 | 0.007 | 1197 | BRITE |
| 2922 | 58560.855 | 58561.055 | 0.200 | 0.023 | 53 | Variable Stars South (Blane) |
| 2926 | 58584.849 | 58585.049 | 0.200 | 0.016 | 150 | Variable Stars South (Blackford) |
| 2928 | 58596.846 | 58597.066 | 0.220 | 0.012 | 35 | Variable Stars South (Blane) |
| 2933 | 58626.839 | 58627.059 | 0.220 | 0.010 | 33 | Variable Stars South (Blane) |

QZ Car was intensely monitored over two successive cycles from La Silla Observatory in 1992 (Mayer 2001). The light curve (Figure 10) shows a puzzling feature in the later cycle which was not obvious in the preceding cycle. Similar, though less dramatic, short-term fluctuations have been recorded by other observers.

More recently the BRITE light curves from 2017 and 2018 showed long term oscillations of minimum and maximum magnitudes of the order of several percent (cf. Figure 1) and the depth of primary eclipses also varied significantly from cycle to cycle (Figure 11).

We initially thought some of the small variations may have been associated with aspect variations in the orbit of the 20.73-day pair. BRITE data were searched for such a signature. There were indications of this but at such a low level that it was not considered significant and not a source of the observed variations.

The source of these variations is unclear but may be related to wind-wind interactions, accretion disc/hot spots due to mass transfer in the eclipsing pair, and/or intrinsic variability of the bright supergiant member of the system.

## 8. Conclusions

Previously published and new observations combined with unpublished measures during the interval 1974 to 1978 have allowed a much more complete O–C diagram of the light-time effects. The six additional early epochs define the LTE curve during most of the period when the eclipsing secondary pair was at its closest to us. This pair is presently approaching again but as yet there is no indication that this approach is slowing to parallel the earlier part of the O–C curve. We conclude that the two pairs of stars in the system have not yet completed one orbit since discovery in 1971.

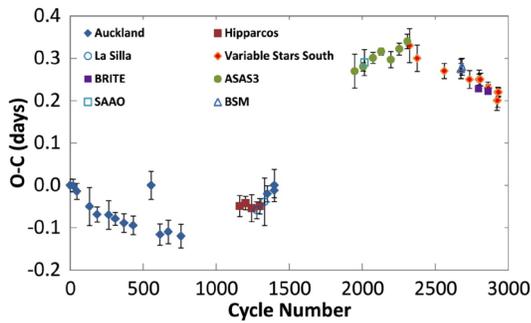

Figure 6. O–C diagram showing light-time effects due to the long-period orbit of the two pairs about each other.

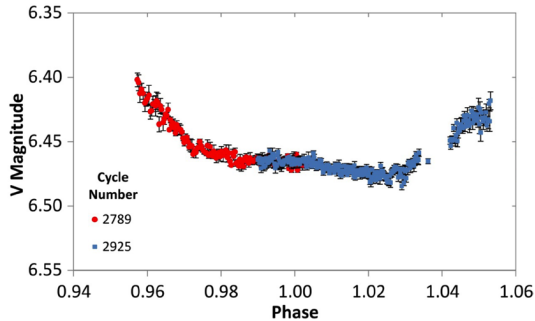

Figure 7. LTE and zero point corrected light curve of primary eclipse composed from ingress measures in 2017 and egress measures in 2019. These cover a period of ~0.57 days or 13.7 hours. The gap on egress was caused by a permanent observing obstruction which could not be removed.

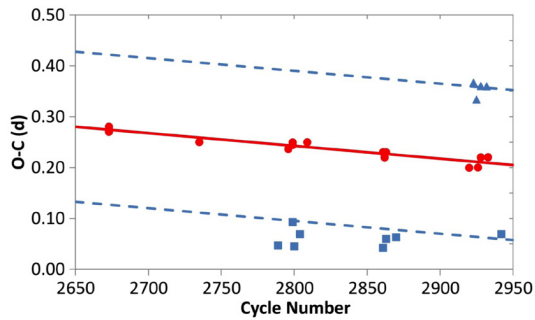

Figure 8: O–C diagram of measured seasonal epochs (red circles) from 2015 through 2019. Also plotted are O–C values determined for the start (blue squares) and end (blue triangles) of totality for primary eclipses. Dashed blue lines indicate totality lasting 0.295 day.

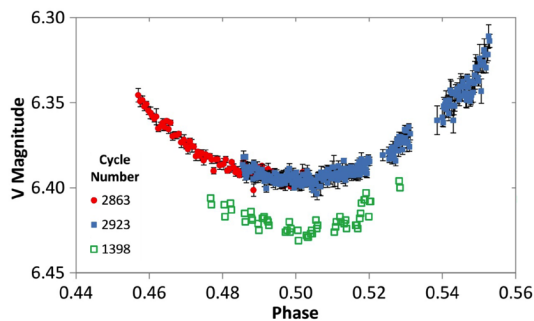

Figure 9. LTE and zero point corrected composite secondary eclipse light curve from ingress measures by MB (red circles) and egress measures by GB (blue squares). Christie's measures are plotted as empty green squares.

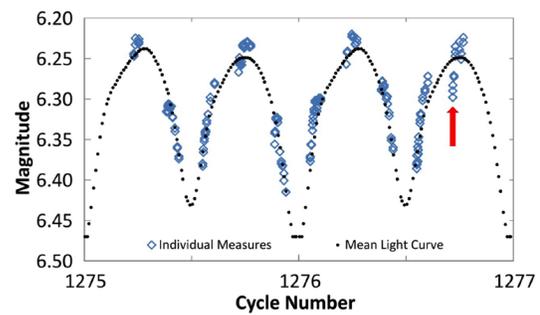

Figure 10. Two consecutive eclipse cycles measured at La Silla Observatory in 1992. Measures during cycle 1275 match the model light curve reasonably well. However, cycle 1276 shows a significant deviation (red arrow) from the model near the maximum following secondary eclipse.

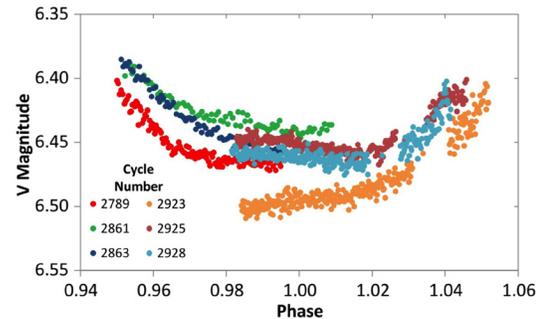

Figure 11. Three sets of primary eclipse ingress measures by MB and three of egress by GB showing significant cycle to cycle brightness variations. The downward slope during totality was consistently observed apart from cycle 2923 which showed an upward slope.

BRITE satellite measures defined the true shape of the light curve which models must be able to emulate. Combined with our measures of primary and secondary eclipse shapes, we were able to produce a mean light curve to which random measures in each season could be fitted to produce reliable seasonal epochs.

Primary eclipse totality was found to last $7.08 \pm 0.48$ hours, which helps to explain why attempts to determine epochs from measures at this phase have been largely unsuccessful. The various figures above illustrate other aspects of the problems quite well.

QZ Car in the open cluster Collinder 228 is one of the brightest objects near the eta Carinae region and its radiation is responsible for stimulating the emission in much of the nebulosity in the southern part of that region. The system's mass and complexity are still not fully understood and continued study is extremely important.

With this in mind ingress in the 2020 season will best be studied from South America and egress from Eastern Australia. We should see the completion of one long period orbit within the next few years. Until then these epochs may help other researchers.

## 9. Acknowledgements

Pavel Mayer and Stan Walker exchanged data on several occasions. This included those labelled in the O–C diagram as La Silla, SAAO, and the secondary eclipse measures of Christie used in Figure 9.

This paper is based in part on data collected by the BRITE Constellation satellite mission, designed, built, launched, operated, and supported by the Austrian Research Promotion Agency (FFG), the University of Vienna, the Technical University of Graz, the University of Innsbruck, the Canadian Space Agency (CSA), the University of Toronto Institute for Aerospace Studies (UTIAS), the Foundation for Polish Science and Technology (FNiTP MNiSW), and National Science Centre (NCN).

APi acknowledges support from the NCN grant No. 2016/21/B/ST9/01126.

APo was supported by Silesian University Rector Grant no. 02/140/RGJ20/0001.

GW acknowledges support in the form of a Discovery Grant from NSERC, Canada.

We acknowledge the AAVSO for the use of AAVSOnet data and appreciate the hard work and dedication of AAVSOnet site managers Peter Nelson (BSM South) and Greg Bolt (BSM Berry).

This research has made use of the database of the All Sky Automated Survey ASAS, http://www.astrouw.edu.pl/asas/.

Members of the Auckland Astronomical Society helped with many of the observations as the Auckland Photoelectric Observers' Group, using the 50-cm Edith Winstone Blackwell telescope until epoch 759, later the Milton Road observatory with a 53-cm Cassegrain for epochs 1350 and 1398.